 \documentclass[final,1p,times]{elsarticle}

 \usepackage{graphicx}

\usepackage{amssymb}

\usepackage{lineno}

\journal{Planetary and Space Science}

\begin{document}

\begin{frontmatter}





\title{The Cratering History of Asteroid (21) Lutetia}

%
\author[mar]{S. Marchi\corref{cor1}}
\ead{marchi@oca.eu}
\author[mas]{M. Massironi}
\author[hol]{J.-B. Vincent}
\author[mar]{A. Morbidelli} 
\author[mot]{S. Mottola}    
\author[marz]{F. Marzari} 
\author[kue]{M. K\"uppers} 
\author[bes]{S. Besse}
\author[nic]{N. Thomas} 
\author[bar]{C. Barbieri}
\author[nal]{G. Naletto}
\author[hol]{H. Sierks}

%
%
\cortext[cor1]{Corresponding author}

\address[mar]{Departement Cassiop\'{e}e,  Universite de Nice  - Sophia
  Antipolis, Observatoire  de la C\^{o}te d'Azur,  CNRS, Nice, France}
\address[mas]{Department of Geosciences, Padova University, Italy}
\address[hol]{Max Planck Institute for Solar System Research, Lindau, Germany}
\address[mot]{Institut f\"ur Planetenforschung, DLR-Berlin, Germany}
\address[marz]{Department of Physics, Padova University, Italy}
\address[kue]{ESA-ESAC, Villanueva de la Ca\~nada (Madrid), Spain}
\address[bes]{Laboratoire d'Astrophysique de Marseille, France}
\address[nic]{Physikalisches Institut, University of Bern, Switzerland}
\address[bar]{Department of Astronomy, Padova University, Italy}
\address[nal]{Department of Information Engineering, Padova University, Italy}

\begin{abstract}

The European Space Agency's Rosetta spacecraft passed by the main belt
asteroid (21) Lutetia the 10$^{\rm th}$ July 2010. With its
$\sim100$~km size, Lutetia is one of the largest asteroids ever imaged
by a spacecraft.  During the flyby, the on-board OSIRIS imaging system
acquired spectacular images of Lutetia's northern hemisphere revealing
a complex surface scarred by numerous impact craters, reaching the
maximum dimension of about 55~km.\\

In this paper, we assess the cratering history of the asteroid. For
this purpose, we apply current models describing the formation and
evolution of main belt asteroids, that provide the rate and velocity
distributions of impactors. These models, coupled with appropriate
crater scaling laws, allow us to interpret the observed crater
size-frequency distribution (SFD) and  constrain the cratering
history.  Thanks to this approach, we derive the crater retention age
of several regions on Lutetia, namely the time lapsed since their
formation or global surface reset. We also investigate the influence
of various factors -like Lutetia's bulk structure and crater
obliteration- on the observed crater SFDs and the estimated surface
ages.\\

From our analysis, it emerges that Lutetia underwent a complex
collisional evolution, involving major local resurfacing events till
recent times.  The difference in crater density between the youngest
and oldest recognized units implies a difference in age of more than a
factor of 10.  The youngest unit (Beatica) has an estimated age of
tens to hundreds of Myr, while the oldest one (Achaia) formed during a
period when the bombardment of asteroids was more intense than the
current one, presumably around 3.6~Gyr ago or older.

\end{abstract}


\end{frontmatter}



\section{Introduction}

The European Space Agency's (ESA) Rosetta spacecraft passed by the
main belt asteroid (21) Lutetia with a relative velocity of $\sim
15$~km/s on 10 July 2010 at 15:44:56~UTC.  The Rosetta-Lutetia
distance at closest approach (CA) was 3170~km.  During the flyby the
solar phase angle (sun-object-observer) decreased from the initial
11$^{\circ}$ to a minimum of 0.15$^{\circ}$ 18 minutes before CA, then
increased again to 80$^{\circ}$ at CA and finally reached a maximum of
139$^{\circ}$ when the observations were stopped.  A total of 400
images were obtained by the Optical, Spectroscopic, and Infrared
Remote Imaging System (OSIRIS), which consists of two imagers: the
Wide Angle Camera (WAC) and the Narrow Angle Camera (NAC)
\citep{kel07}.  The best resolution at CA corresponded to a scale of
60~m/px at the asteroid surface.\\
Lutetia has an orbital semi-major axis of about 2.43~AU, an
eccentricity of 0.16 and an inclination of 3.06$^{\circ}$.  Its shape
can be fitted by an ellipsoid having axes of $121 \times 101 \times
75$~km \citep{sie11}.\\
 
Previous space missions have visited and acquired detailed data for a
total of 6 asteroids, namely four main belt asteroids \citep[951
  Gaspra, 243 Ida, 253 Mathilde, 2867
  Steins;][]{vev99a,bel92,bel94,kel10} and two near-Earth objects
\citep[433 Eros, 25143 Itokawa;][]{vev99b,sai06}.  Itokawa is the
smallest of them, with dimensions of $0.45 \times 0.29 \times
0.21$~km.  The other asteroids have average sizes ranging from
$\sim5$~km to $\sim53$~km.  In this respect, Lutetia with its average
size of 98~km is the second largest asteroid ever visited by a
spacecraft so far (at the moment of the writing -October 2011-, Dawn
mission is orbiting around the 500-km sized asteroid  (4) Vesta).\\

This paper analyzes some of the highest resolution OSIRIS images with
the aim to study the crater size-frequency distributions (SFDs) on the
different units that have been identified on the basis of geological
investigations \citep{sie11,mas11,nic11}.  This analysis provides
constraints on Lutetia's bulk structure and surface evolution.  The
observed crater SFDs are also used to compare the cratering process
among the different units, to derive absolute ages and provide a
chronology of the major events that affected Lutetia evolution.\\

\section{Lutetia crater population}

\begin{figure*}[h] 
\includegraphics[width=10cm]{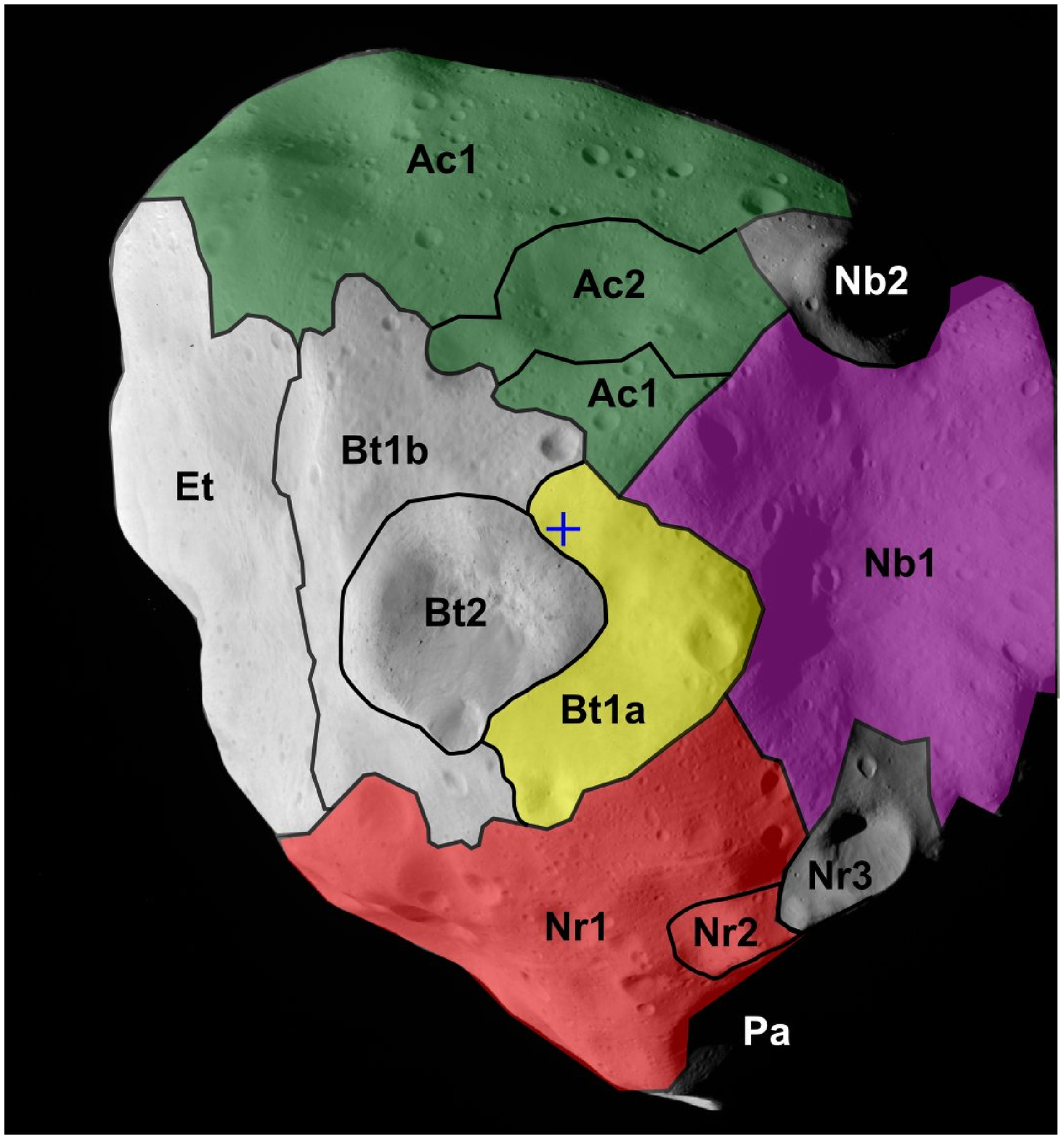}
\caption{ The 6 major regions identified on Lutetia.  Ac: Achaia, Nr:
  Noricum, Nb: Narbonensis, Bt: Baetica, Et: Etruria, Pa: Pannonia.
  Note that some bondaries may sligtlhy vary according to different
  authors. For a more detailed definition of the regions see
  \cite{mas11} and \cite{nic11}.  Some of the major units
  (i.e. subdivisions of the regions) are also reported.  Colored units
  are those used for crater counts. The corresponding areas (km$^2$)
  are: 760 (Bt1a), 2875 (Ac1+Ac2), 2042 (Nr1+Nr2), 2647 (Nb1).  The
  blue ``+'' at the center of the image indicates the north pole.}
\label{units}
\end{figure*}

The NAC high resolution images acquired during the flyby where used to
identify major regions on Lutetia (see Fig. \ref{units}). These
regions have been defined by taking into account several factors,
including local topography, geological features, surface texture,
crater spatial density and stratigraphic relationships
\citep{nic11,mas11}.  In this respect, each region is characterized by
distinct properties of one or more of the above listed factors.  The
regions indentified have been further divided into several units.
Thanks to this selection criterion, the defined units reflect major
differences in their evolution \citep{mas11,nic11}.  Note that the
actual unit boundaries are in some cases not well established due to
the lack of resolution and/or unfavorable illumination conditions
\citep{mas11,nic11}.\\

Among the major regions, only 4 were imaged with enough quality for
accurate crater counting to be performed.  These are Achaia,
Narbonensis, Noricum and Baetica.  Their geological properties show
remarkable differences, therefore they will described individually in
the following sections.\\
 
{\bf Achaia.} This region is defined by a remarkably flat and uniform
area. It is bounded by Baetica, Narbonensis and Etruria. Its
boundaries with Baetica and Narbonensis are defined mainly by texture
and topography, respectively. The boundary with Etruria is defined by
the same means but, due to low contrast of the images in these
regions, it is less precisely established\footnote{Note that several
  choices of the Etruria-Achaia boundary have been performed in 
  our analysis. The influence on the actual choice on the resulting
  Achaia crater SFD is negligible.}.  The illumination conditions
within Achaia are very good and uniform, therefore craters are clearly
visible and their size estimate is performed with precision
\citep{vin11}.\\

\begin{figure*}[h] 
\includegraphics[width=11cm]{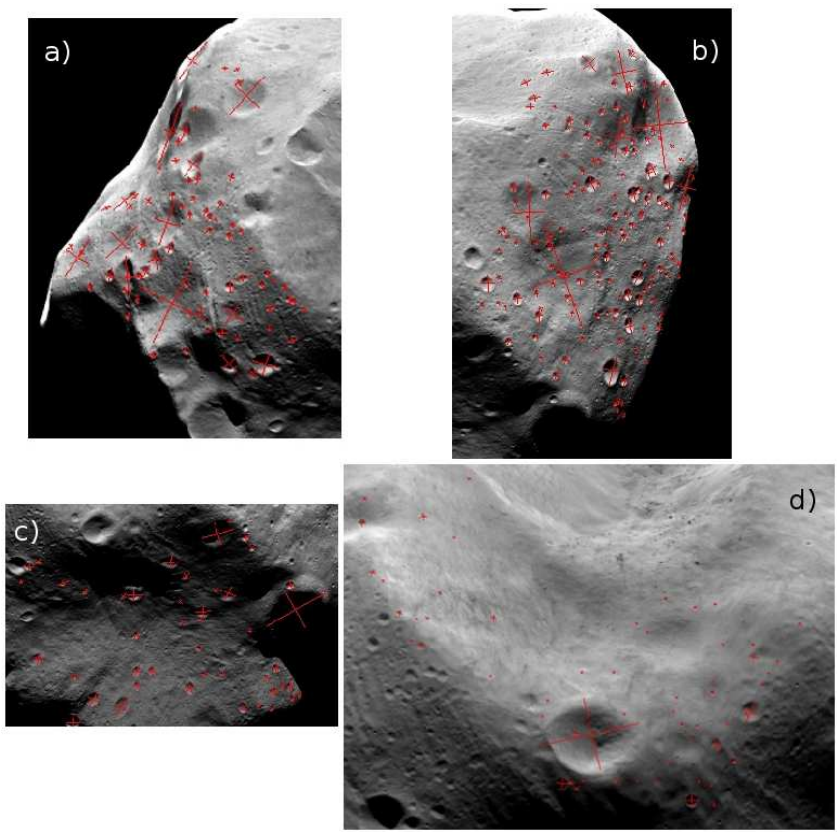}
\caption{The panels indicate the craters counted on the 4 units
  investigated. Crater counts have been performed on image
  NAC.15.42.41.240 for Achaia (b), Noricum (a) and Narbonensis (c)
  regions, and on image NAC.15.44.41.262 for Bt1a region (d).}
\label{craters}
\end{figure*}

The Achaia region (Ac1+Ac2) is heavily cratered, showing a large range
of crater sizes, from 21.6~km (Nicea crater) down to the resolution
limit (we used a minimum of 4~pixels to identify craters, thus about
0.2~km). The overall spatial distribution of the 157 craters $>0.6$~km
is uniform and there appears to be no evident contamination from
adjacent units (see Fig. \ref{craters}, panel b).  At smaller sizes,
several crater-like features may not be of impact origin. Many
circular depressions are close to, or overlap linear features,
therefore may not represent bona fide craters.  The presence of
secondary craters (formed by boulders ejected during the formation of
other craters) can also be possible at these small crater sizes,
although it is unclear how likely can secondary craters form on Lutetia,
given its low escape velocity.\\

\begin{figure*}[h] 
\includegraphics[width=9cm,angle=-90]{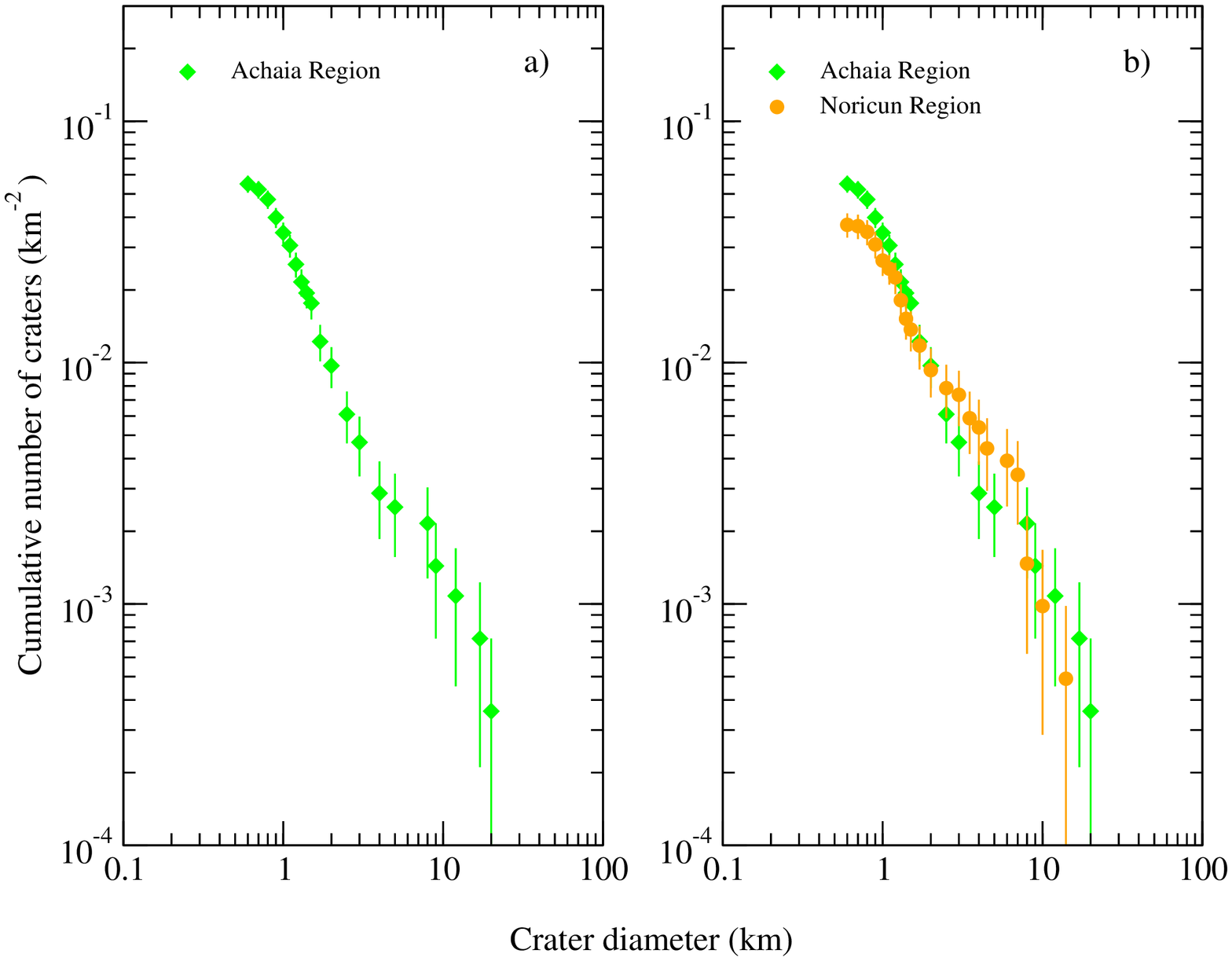}
\includegraphics[width=9cm,angle=-90]{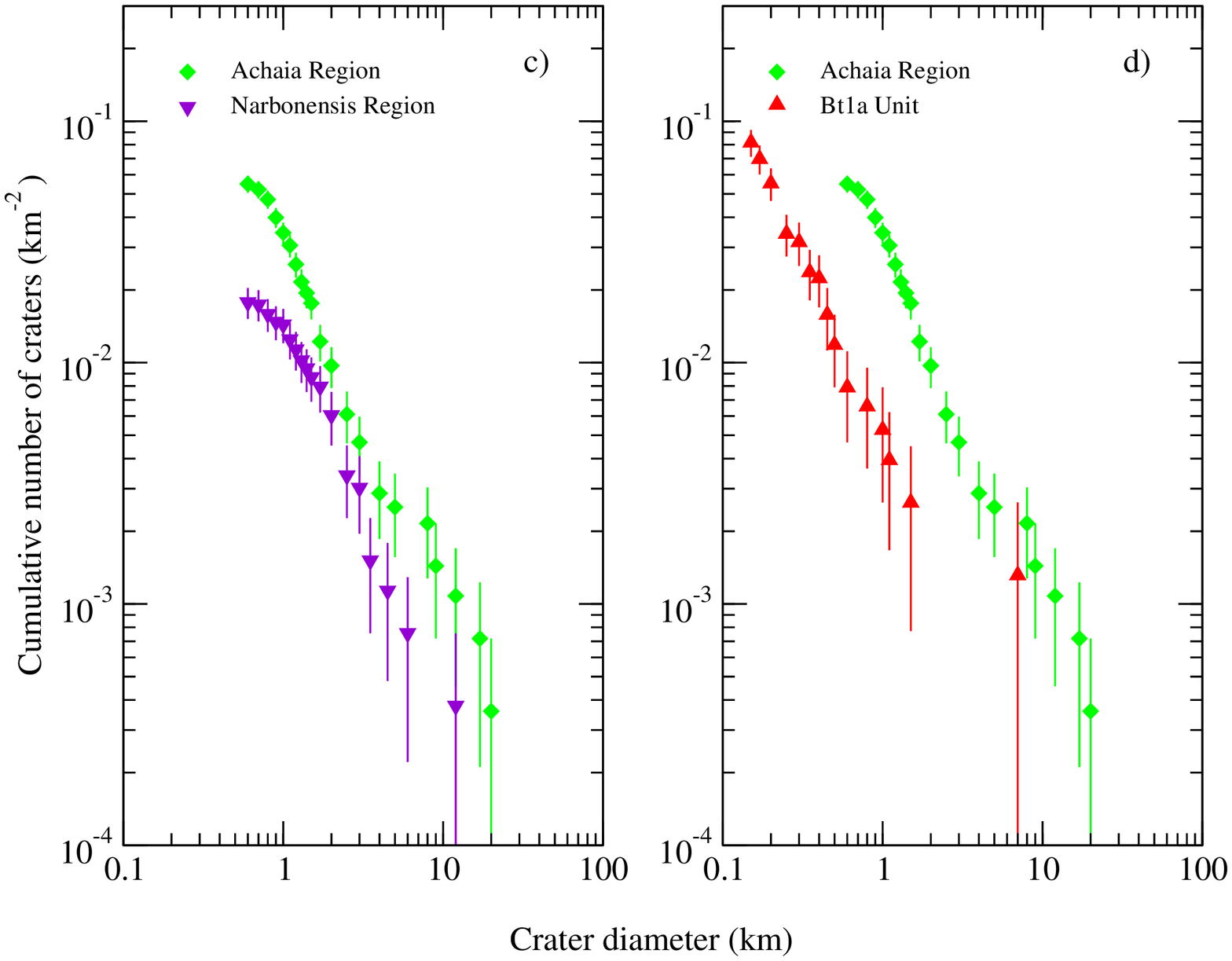}
\caption{Crater size-frequency distributions of the 4 units reported
  in Figure~\ref{craters} (Achaia region (a), Noricum region (b),
  Narbonensis region (c), Bt1a unit (d)).  Achaia crater SFD is
  reported in all panels for a better comparison.}
\label{csfd}
\end{figure*}

For the purpose of age assessment, we are interested in primary craters
(i.e. formed by impacts with asteroids), therefore our analysis
focuses on craters $>0.6$~km. The resulting crater SFD is shown in
Fig. \ref{csfd} (panel a).\\

An interesting result is that Achaia's crater SFD exhibits a marked
flexure point at about $4-7$~km.  Note that the observed flexure point
is unlikely due to observational biases, like uncertainties in the
identification of craters or resolution issues. This is because Achaia
region is a remarkably flat area and it has been imaged with uniform
conditions of illumination, while the flexure point is well above the
image resolution. Moreover, thanks to the boundary selection, we also
exclude that the observed flexure is due to obliteration of small
craters due to crater ejecta coming from nearby units (e.g.,
Beatica). For the same reason, it seems also unlikely that the
  formation of the large crater Massalia (see next sections) played a
  role in the formation of the flexure point in Achaia crater SFD.\\

{\bf Noricum.}  This unit has a very complex topography. It contains a
number of closely packed and prominent circular features, likely
impact craters, showing several stages of degradation \citep{vin11}.
Moreover, this unit looks ``compressed'' among the impact craters of
Baetica, Massalia crater, and possibly another large crater on the
dark side of Lutetia (namely, Pannonia region; see Fig.~\ref{units}),
the presence of which may be inferred thanks to the circular
terminator of part of Noricum.  These factors are likely at the origin
of Noricum complex topography.\\

Crater counts have been performed in unit Nr1+Nr2 (for semplicity we
will refer to Noricum region in the rest of the work).  The overall
viewing geometry is not optimal (i.e. nearly edge-on), therefore the
size estimate of some of the 76 identified craters ($>0.6$~km) is
problematic (see Fig.~\ref{craters}, panel a).  The resulting crater
SFD shows a clear transition at about 2~km (see Fig.~\ref{csfd}, panel
b): the slope of the crater SFD for $D>2$~km is considerably shallower
than that for $D<2$~km. The feature resembles somewhat the flexure seen
on Achaia crater SFD, although in this case it may be due to imprecise
size estimate for several large craters due to their nearly edge-on
view. The crater spatial density for $D<2$~km is very similar to that
of Achaia.\\

\begin{figure*}[h] 
\includegraphics[width=10cm]{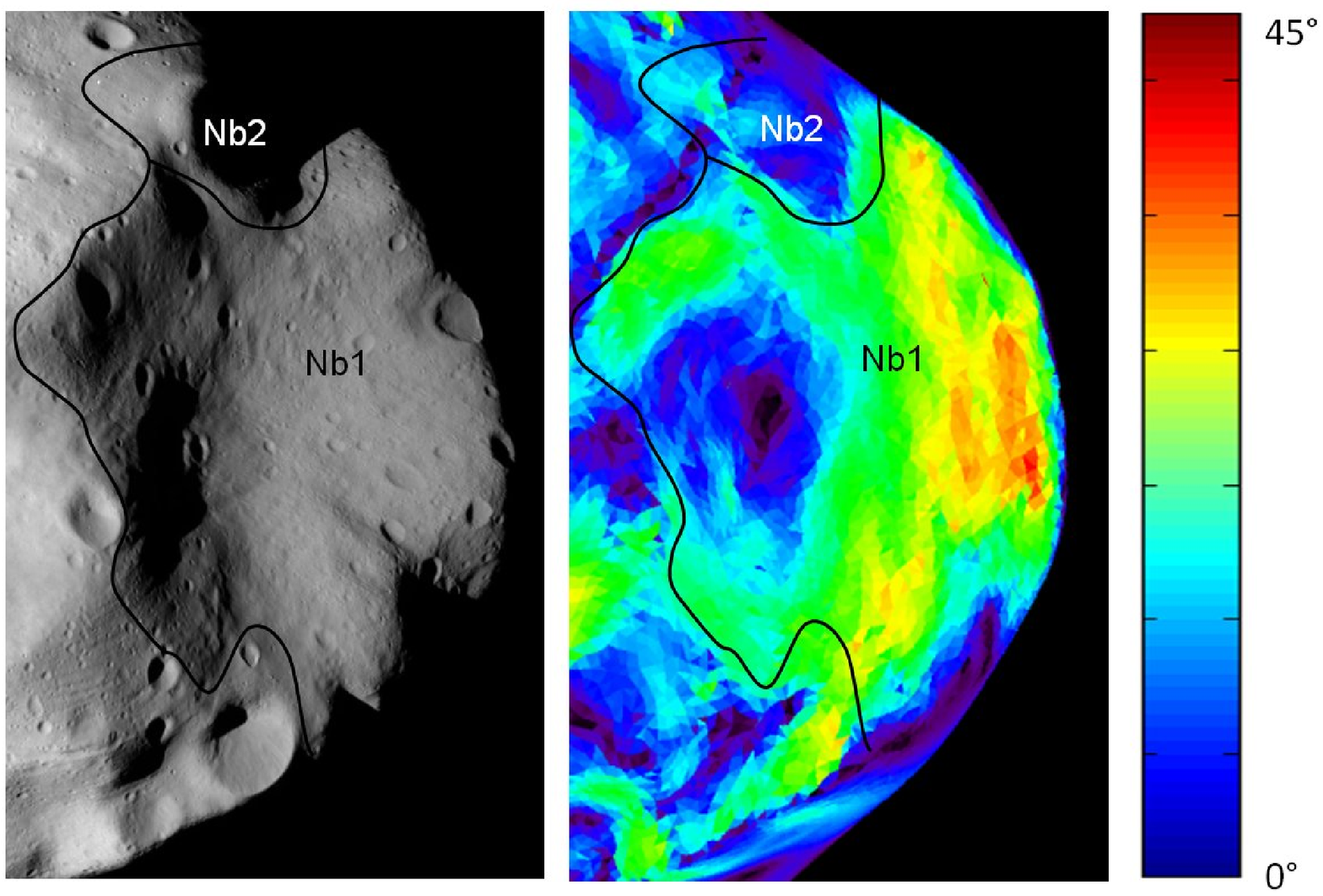}
\includegraphics[width=10cm]{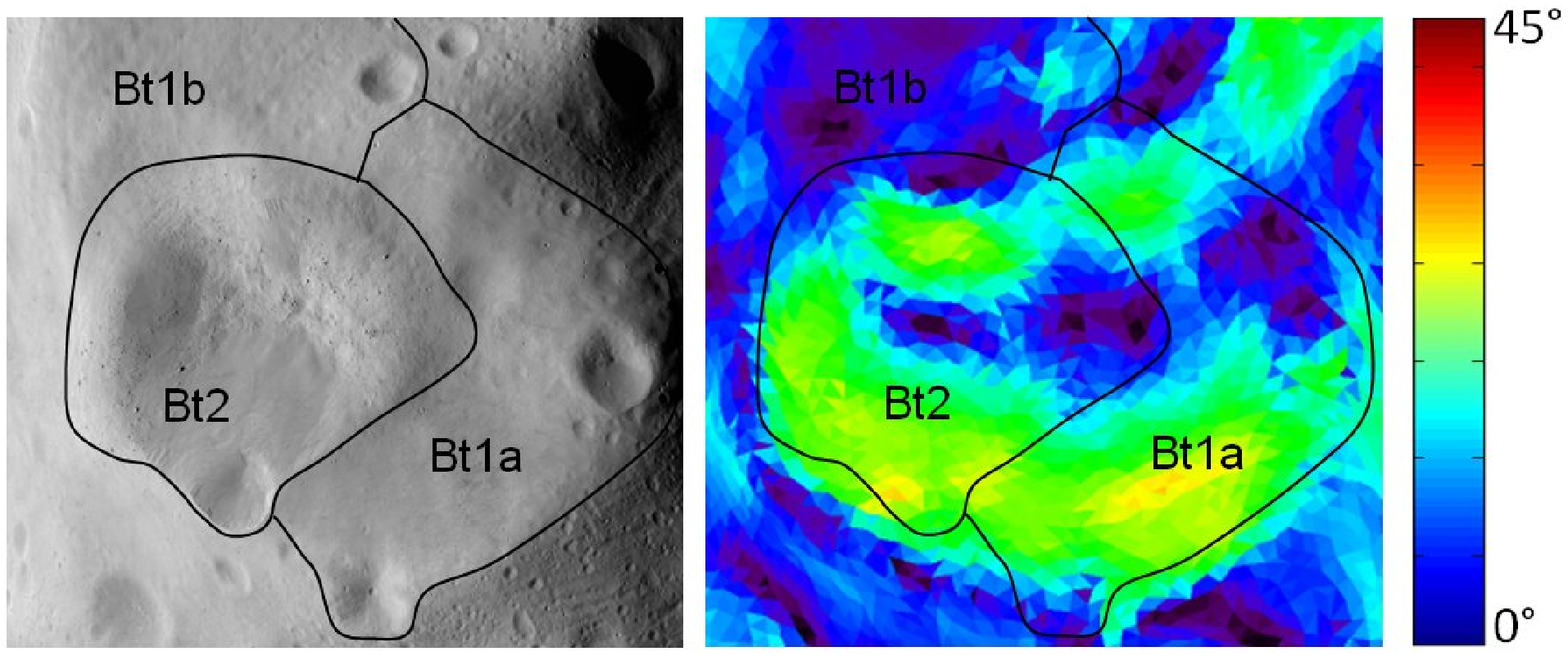}
\caption{Close view of Narbonensis and Baetica regions (top-left and
  bottom-left panels, respectively). Topographical slope (i.e., the
  angle between local shape and local gravity) for Narbonensis
  (top-right panel) and Baetica (bottom-right panel) regions.}
\label{slopes}
\end{figure*}

{\bf Narbonensis.} This region corresponds to the interior of the
55-km-sized crater Massalia, the largest impact structure detected on
Lutetia. Crater count has been performed in unit Nb1 (for simplicity
we will refer to Narbonensis region in the rest of the work).  A total
of 47 craters $>0.6$~km have been identified (see Fig. \ref{craters},
panel c). Notably, several craters appear deformed by sliding of their
rims due to the relatively high topographic slope present in large
part of the unit \citep[][see also Fig.~\ref{slopes}, upper panels]{vin11}. 
In these cases, the determination of the actual
crater size is not very accurate.\\

Overall, the crater spatial density of Narbonensis is lower than that
of Achaia (see Fig. \ref{csfd}, panel c). The shapes of the crater
SFDs of the two units also differ. In particular, the Narbonensis
crater SFD has a shallower slope at small sizes than Achaia. It is not
clear whether this difference is due to poor count statistics or it is
a real feature. In the latter case, it might be due to variation in
the local properties of the terrains or due to some later modification
(as we will discuss later).\\

{\bf Baetica.} This region, unlike the previous ones, shows marked
evidence of several major modification processes (landslides, ejecta
blanketing etc) that have been used to establish sub-units that likely
formed at different epochs \citep{nic11,mas11}.  Moreover, this region
is also characterized by large topographical slope variations (from 0
to 45~deg, see Fig.~\ref{slopes}, lower panels), and by the presence of
many large boulders \citep{kue11}.\\

Overall, the Baetica region presents much fewer craters than adjacent
regions. Some Baetica's units appear extremely young, showing no
detectable impact craters. For these reasons, we restrict our analysis
to a unit, named Bt1a (see Fig.~\ref{units}), which apparently has not
been affected by recent geological processes (e.g., landslides), it is
relatively flat and uniform, and does contain a fair number of small
impact craters.  In this case, we boost crater detection by using
Laplacian-filtered images\footnote{ The Laplacian filter technique uses
  secondary derivatives in two directions to enhance the contrast of
  the input image and it is known to be very effective in revealing
  small, high frequency features \citep{bes11}. }.  We identify 62
craters in the range $0.2-1$~km (see Fig.~\ref{craters}, panel d).\\

The Bt1a crater SFD shows an overall shape consistent with those of
other units, and it is characterized by a much lower crater spatial
density (see Fig. \ref{csfd}, panel d).  Interestingly, Bt1a contains
a fresh and large ($\sim7$~km) crater, plus a second highly degraded
crater having similar dimensions that has not been counted since it
probably formed before Bt1a \citep{nic11,vin11}.\\

\section{The Model Production Function chronology}

The crater SFDs of the units presented in the previous section can be
used to derive their crater retention ages.  The age of units is
crucial information, since it provides constraints on the formation
and evolution of Lutetia. In this respect, Lutetia stands out with
respect to all previously visited asteroids (except Vesta), for its
complex geological evolution. Therefore, the crater retention ages of
its units are important to set a timeline for this evolution.
Moreover, the study of the cratering process along with geological
assessment can be used to constrain the physical properties of the
target.\\
 
In this work, crater retention ages are derived in the framework of
the Model Production Function (MPF) chronology \citep{mar09}.  With
this approach Lutetia's crater production function (i.e., the expected
number of craters per year per unit surface) is computed by modeling
its impactor flux and by using a crater scaling law in order to
compute the resulting crater population. The resulting crater MPF
  gives the cumulative density of craters (per year) as a function of
  the crater size.\\
 
\begin{figure*}[h] 
\includegraphics[width=11cm,angle=-90]{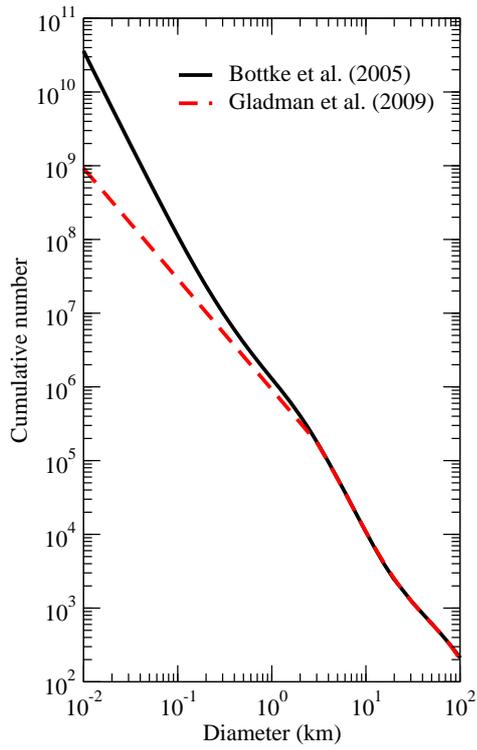}
\caption{Main belt size-frequency distributions used in this work.
  The model \cite{bot05a} distribution is based on the observed
  debiased main belt population down to about 1~km. The \cite{gla09}
  distribution is obtained in the following manner: for impactor sizes
  $\ge3$~km (corresponding to the completeness limit of $H$-magnitute
  $\sim15$, for an assumed albedo of 0.2), it overlaps with the
  \cite{bot05a} SFD; while for sizes $<3$~km it is has a cumulative
  slope of -1.5. Note that the SKADS survey is valid down to $H\sim18$
  (correponding to about 0.8~km), nevertheless we extrapolated the
  -1.5 slope to smaller sizes.}
\label{mba}
\end{figure*}

\begin{figure*}[h] 
\includegraphics[width=11cm,angle=-90]{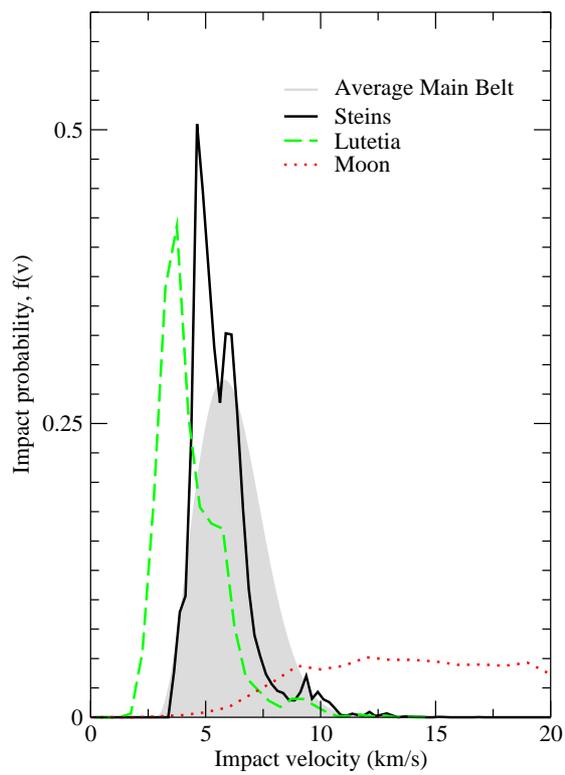}
\caption{Lutetia's impact velocity distribution. For a comparison, the
  distribution for Steins, the average main belt (shaded area) and the
  Moon (largely out of scale) are also reported. The average impact
  velocity for Lutetia is 4.3~km/s.}
\label{vel}
\end{figure*}

In analogy with previous work, the impactor flux is characterized by
its size-frequency distribution and impact velocity distribution.  The
impactor SFD is taken from the model population of main belt asteroids
of \cite{bot05a}.  In this work, we will also consider a second
  MBA population derived by the Sub-Kilometer Asteroid Diameter Survey
  (SKADS) \citep[][ see Fig.~\ref{mba}]{gla09}. Using the
  \cite{far92} algorithm, we computed that the intrinsic collision
  probability between MBAs and Lutetia is
  $P_i=4.21\cdot10^{-18}$~km$^{-2}$yr$^{-1}$. Note that this $P_i$
  value is significantly higher that the average value for the main
  belt, namely $2.86\cdot10^{-18}$~km$^{-2}$yr$^{-1}$. Using the same
  algorithm, we also computed the Lutetia's impact velocity
  distribution (see Figure~\ref{vel}).\\

Concerning the crater scaling law, we adopted a Pi-group scaling law
\citep{hol07}. These scaling laws allow us to estimate the size of a
crater given the dimension ($d$) and velocity ($v$) and density
($\delta$) of the impactor along with the density ($\rho$) and
strength ($Y$) of the target. In addition to these quantities, two
parameters ($\nu, \mu$) account for the nature of the terrains
(hard rock, cohesive soil, porous material).  In this paper, we
investigate both hard rock and cohesive soils scaling laws, whose
parameters are $\nu=0.4, \mu=0.55$ and $\nu=0.4, \mu=0.41$,
respectively.  We assume $Y=2\cdot10^8$~dyne/cm$^2$ for typical
hard rock and an impactor density of $\delta=2.6$~g/cm$^3$
\citep{mar10}.  The bulk density of Lutetia is $\rho=3.4$~g/cm$^3$
\citep{sie11}.  Values of density and strength for cohesive soils will
be given in Section~4. Further details about the crater scaling law
can be found in \cite{mar11}.  Note that no correction for the
transient-to-final crater size has been applied, because the crater
modification stage is not likely to occur on Lutetia given its low
gravity.\\

Absolute ages can be computed by knowing the time dependence of the
impactor flux in the past. Unfortunately, such time-dependence is not
known for main belt asteroids.  Two approaches can be used to overcome
such a limitation.  First, one can assume that the present impact rate
for main belt asteroids remained constant over the age of the solar
system. This scenario requires a constant main belt population, where
no big modification (e.g., in its orbital architecture and total mass)
occured. However, it is known that the main belt was more massive in
the past and that during the early phases of the solar system it was
shaped by major events \citep[e.g.,][]{mor10}.  However, these
processes have not yet been modeled with enough certainty and accuracy
to enable the determination of the time evolution of the impact rate.
An alternative approach is to refer to the lunar impactor flux, which
has been calibraterd on the basis of radiometric ages of lunar samples
\citep{neu94,mar09}.  This scenario assumes that the impactor flux
variation experienced by the Moon also applies to main belt asteroids.
In reality, since the Moon is not embedded in the main belt, it is
likely that the Moon and MBAs had very different impact histories.
For instance, consider the case that the lunar impact cataclysm
between 4.1 and 3.8~Gy ago was due to a temporary destabilization of
the main belt that removed a part of its asteroids. Then the Moon
would have suffered an impact spike, while the impact rate in the
asteroid belt would have decreased (i.e., without impact spike) from
an initial higher but rougly constant value in the $4.5-4.1$~Ga
time-interval, to the current value. Thus, the time evolutions of the
impact rate on the Moon and in the asteroid belt would have been
totally different.  On the other hand, in the case of a large cometary
contribution to the lunar cataclysm (i.e., from a source region
outside the main belt) these bodies would have produced an impact
spike on both the Moon and MBAs.\\

In the assumption that the evolution of the impact rate in the
asteroid belt and on the Moon was the same, and assuming that all
craters that are formed are retained on the surface, the crater MPF
function for an asteroid at a time $t$ is given by:

\begin{equation}
{\rm MPF}(D,t)={\rm MPF}(D,1{\rm yr})\cdot\frac{N_1(t)}{N_1(1{\rm yr})} \label{mpf_eq}
\end{equation}

where $D$ is the crater size and $N_1(t)$ expresses the lunar crater
cumulative number at 1~km as a function of time according to the
following equation:

\begin{equation}
N_1(t)=a(e^{bt}-1)+ct \label{chrono_eq}
\end{equation}

where $t$ is in Gyr ($t=0$ is the present time),
$a=1.23\times10^{-15}$, $b=7.85$, $c=1.30\times10^{-3}$
\citep{mar09}\footnote{The parameters $a,b,c$ for the lunar
  chronology curve are determined by best fit of lunar calibration data
  and their actual values may vary according to different authors
  \cite[e.g.,][]{neu94}.  However, the variation of the actual values
  for $a,b,c$ has a negligible impact on the age determination. }.
Note that setting $a=0$ would correspond to the constant flux
scenario.  The MPF($D,t$) is used to derive the model cratering age by
a best fit of the observed crater SFD that minimizes the reduced chi
squared value, $\chi_r^2$.  Data points are weighted according to
their measurement errors.  The formal errors on the best age
correspond to a 50\% increase of the $\chi_r^2$ around the minimum
value. Other sources of uncertainties are neglected \citep[see][for
  more details]{mar11}.\\

Equation~\ref{mpf_eq} basically implies that MPF($D,t$) is obtained by
simply y-axis shifting MPF($D$,1yr) by a proper amount.  It has been
shown by previous studies on asteroid cratering, however, that several
crater obliteration processes may be at work
\citep[e.g.,][]{obr06}. In the case that crater obliteration occurs,
the shape of the MPF changes over time and may reach a steady-state in
the case that crater saturation occurs (namely, the newly formed
craters erase previous ones leaving the overall crater spatial
density unchanged).  In this paper, we take into account crater
obliteration processes as described in \cite{mar10}.\\

\section{Crater retention age estimates}

One important aspect of MPF methodology is that it depends on the
assumed properties of the target body \citep{mas09,mar11}. Therefore,
the analysis of crater SFDs on different terrains on the same body (or
different asteroids) should be done with caution, since changes in the
material properties may invalidate direct comparison \citep{mar11}.
Generally speaking, material properties are not known in detail,
however, in some cases, they can be constrained on the basis of
geomorphological and geological analysis.  Therefore, whenever
possible, MPF chronology allows to derive cratering ages taking into
account explicitly the effect of the inferred material properties. In
this section we present the results of our MPF-based age estimate for
each unit investigated.\\

\begin{figure*}[h]
\includegraphics[width=11cm,angle=-90]{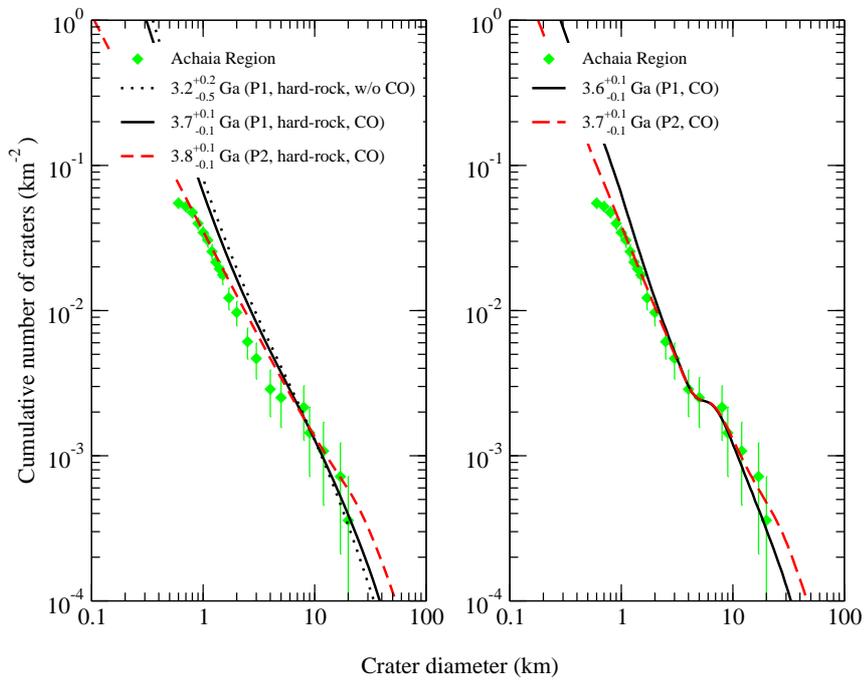}
\caption{Achaia MPF best fit. Left panel: Best fits obtained using
  hard rock scaling law, with and without crater obliteration (CO),
  and using the MBA population from \cite{bot05a} (P1) and \cite{gla09}
  (P2). Right panel: Best fits obtained modeling a transition in the
  physical properties of Achaia region, namely adopting a fractured
  layer onto a more competent interior (see text for more details).}
\label{achaia_mpf}
\end{figure*}

{\bf Achaia.} As described in the previous section, Achaia crater SFD
is characterized by a flexure point located at
$4<D<7$~km. Figure~\ref{achaia_mpf} reports the results of MPF best
fitting of the observed crater SFD.  The left panel shows the best
fits obtained by using \cite{bot05a} population (P1 hereinafter) and
the crater scaling law for hard rock both with and without crater
obliteration. Concerning the crater obliteration process, we took into
account local regolith jolting and crater superposition and adopted
the same parameters used by \cite{obr06}.  Global seismic effects and
cumulative seismic shaking have not been considered because of the
large size of Lutetia\footnote{ To see this, we rescaled the Ida's
    global erasing curve from the Figure~4 of \cite{obr06} to Lutetia.
    It results that a crater of about 100~km would be needed to
    globally erase craters $\ge1$~km. This result also suggests that
    it is unlikely that the formation of Massalia crater triggered
    global surface reset.}.  The present fits are achieved anchoring
the MPF to the large crater end of the crater SFD.  The quality of the
fit is basically the same in the two scenarios, except for a slightly
older age in case of crater obliteration.  These results clearly show
that P1 is not able to accurately reproduce the observed cratering. A
similar conclusion is reached also using the \cite{gla09} population
(P2 hereinafter).  In particular, the observed flexure in the crater
SFD has no correspondence is either MBA populations.  Indeed, the
impactor population is not known at the impactor size relevant for the
flexure ($\sim0.5-0.8$~km) and therefore it is possible that the real
main belt SFD may account for it. Nevertheless, the fact that such a
feature has not been observed on other large asteroids, like Ida and
Mathilde \citep{sie11}, makes this unlikely.\\

We have excluded that the flexure is due to the impactor flux, global
and local obliteration processes, and observational biases (see also
discussion in Section~2).  A further possibility is that the flexure
is related to terrain properties. As shown for Mercury \citep{mar11},
the presence of a stratified target having fractured material at the
surface overlying a more competent interior would produce a crater SFD
showing a characteristic flexure. Such a flexure is the combined
result of i) adopting different material parameters for the fractured
layer and the competent interior and ii) using cohesive soil and hard
rock scaling laws for the two layers \citep{mar11}.  The position of
the flexure is mainly determined by the thickness of the fractured
material, which can be chosen in order to produce a best fit of the
observed crater SFD.  We investigated this possibility, by modeling a
transition in the Achaia properties, as done in \cite{mar11}. The
results are shown in Figure~\ref{achaia_mpf} (right panel). The P1
best fit is now improved, being in overall good agreement with the
crater SFD. The resulting age is $3.6\pm 0.1$~Ga, obtained for a
fractured layer depth of 3~km.   It must be clear that the above
  age derives from the lunar chronology (Equ.~\ref{mpf_eq}), whose
  applicatibily to main belt asteroids is unclear.  It is also
  noteworthy that extrapolating the present main belt impact rate in
  the past would lead to an age older than that of the solar system.
  This suggests that the main belt experienced a heavy bombardment in
  the past, although not necessarily with the time-dependence
  described by equ.~\ref{mpf_eq}. The use of the lunar chronology
  probably provides a lower bound to the real age, whereas the age
  computed assuming a constant flux provides an upper bound (in this
  case a trivial one).\\

The best fit presented in figure~\ref{achaia_mpf} shows a residual
mismatch for craters $0.6-2$~km (much above resolution limit), the
origin of which is unclear. Here we show that, using a shallower MBA
population -such as P2- would produce a better match of the observed
crater SFD.  The resulting age is $3.7\pm 0.1$~Ga.  Note, however,
even in the presence of a shallower population a stratified target is
needed in order to explain the flexure (Figure~\ref{achaia_mpf}, left
panel).  It must be clear that the SKAD survey is valid down to an
absolute magnitude of $\sim18$ (corresponding to a size of 0.8~km for
a geometric albedo of 0.15). Such impactors would produce crater sizes
of the order of several km, therefore in our fit we extrapolated P2
slope outside its range of validity.\\

We also find that, independently of the MBA population used, the
Achaia crater SFD is not saturated. Indeed,  at least with the
  crater obliteration parameters adopted here, the saturation occurs
at an higher crater density than observed on Achaia. However, we
caution that this conlcusion depends on the not-well-known process of
crater obliteration. A more thorough analysis of this issue is
deferred to future work.  \\

{\bf Baetica.} This unit is characterized by the presence of a
widespread regolith layer. The thickness of this layer is unknown,
although both crater and landslide morphologies have been used to
constrain its depth to be at least 100s of meters \citep{vin11}.
Therefore, it seems likely that all the craters (except maybe for the
few largest ones) detected in Bt1a formed in highly granular, cohesive
soils.  As for the strength, reference values are from the lunar
regolith ($Y\sim3\cdot10^{4}$ at a depth of 3 meters) and terrestrial
alluvium ($Y\sim7\cdot10^{5}$). Here, we investigate strength values
ranging from $10^5$ to $10^7$~dyne/cm$^2$. We also take a density of
$2$~g/cm$^3$, typical of lunar regolith.\\

\begin{figure*}[h]
\includegraphics[width=11cm,angle=-90]{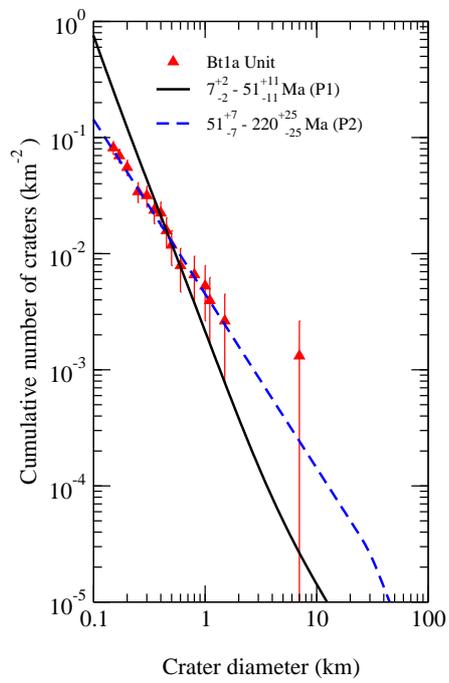}
\caption{Baetica MPF best fit for two impactor populations. The crater
  scaling law for cohesive soil has been used. The derived age ranges
  correspond to two limiting values of strength, namely 10$^5$ and
  10$^7$~dyne/cm$^2$, respectively.  No crater obliteration has been
  applied here, given the very young ages involved (see text for
  further details). }
\label{bt1a_mpf}
\end{figure*}

The resulting MPF best fit, using P1 population, is shown if
Fig.~\ref{bt1a_mpf}.  The main conclusion is that Bt1a is very young,
ranging from $\sim4$ to $\sim50$~Ma, according to the value of the
strength used. The same figure also shows the best fit achieved with
P2. The quality of the fit is now much improved, given the SKADS'
shallower SFD slope. In this case the derived ages range from
$\sim50$ to $\sim220$~Ma. Note that the last age is in better
agreement with the boulder lifetime estimated for the central (and
youngest) unit of Baetica \citep{kue11}.\\

The overall wavy shape of the observed crater SFD is not accurately
reproduced by the MPFs. This may have several explanations, including
low crater statistics and a poor knowledge of the MBA SFD at these
small sizes.  Note that it is also possible, given the large
topographical slopes present in this area, that small craters are not
well preserved (see Fig.~\ref{slopes}, lower panels).\\

{\bf Noricum and Narbonensis.} These two units present several
difficulties in their age assessment. Both crater SFDs are not well
fit by the MPFs, possibly because of errors in the crater size
measurements (Noricum) and poor statistics (Narbonensis).  Some
constraints on the expected evolution of these units come for
geological analysis. First of all, it is clear from stratigraphical
arguments that Massalia crater formed later in time with respect to
both Achaia and Noricum \citep{mas11}.  Therefore, the Narbonensis
unit is younger than Achaia and Noricum.  Moreover, in the light of
the arguments discussed in previous sections, it appears difficult
that the formation of Massalia globally reset Lutetia's surface.  This
conclusion is also in agreement with hydrocode simulations of Massalia
formation. Note that these simulations \citep{cre11} predict that the
Massialia event triggered the formation of a fractured layer generated
all over the surface of Lutetia.  The actual damage of the fractured
layer depends on the resolution of the simulations, nevertheless it is
believed to be not sufficient to cause global resurfacing
(K. W\"unnemann pers.  comm. on 27 June 2011)\footnote{ We also
  acknowledge the fact that these conclusions are based on scaling
  laws and simulations which depend on poorly constrained
  parameters. Thus, it is possible -although unlikely- that the
  formation of Massalia crater triggered major crater reset on nearby
  regions.}.\\

\begin{figure*}[h]
\includegraphics[width=11cm,angle=-90]{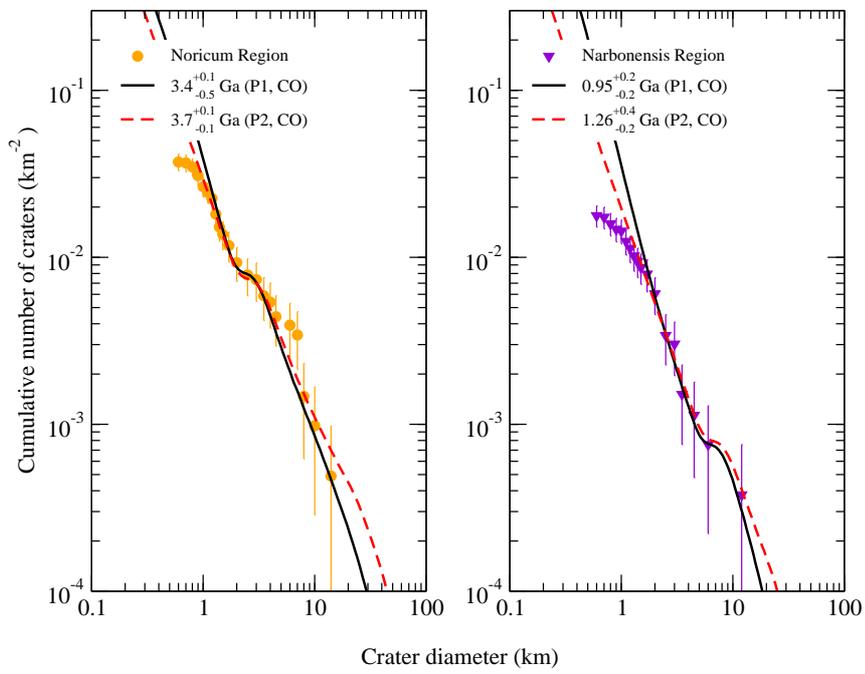}
\caption{Noricum and Narbonensis MPF best fits (see text for further
  details).}
\label{no_nb_mpf}
\end{figure*}

If the above scenario is correct, then we expect that Noricum has
similar properties as Achaia. Figure~\ref{no_nb_mpf} (left panel)
shows the MPF best fit using a stratified target model and crater
obliteration. The best fit is achieved with a fractured layer depth of
1.3~km. The resulting Noricum age is $\sim3.4$ and $\sim3.7$~Ga for P1
and P2 respectively, which is consistent with being coeval with Achaia
(again, we point out that these ages are derived using the lunar
chronology).  The relatively shallow layer of fractured material may
also be consistent with the complex topography of the units (possibly
reflecting a more competent near-surface interior).\\

Figure~\ref{no_nb_mpf} (right panel) shows the MPF best fit of
Narbonensis.  In this case, the observed crater SFD does not show
evidences of a flexure, possibly due to the poor crater statistics,
that would suggest the presence of stratified terrains.  Nevertheless,
according to \cite{cre11}, the interior of the Massalia crater is
expected to be fractured up to the depth of several km.  Therefore, by
using the same crater scaling law and a fractured layer depth of
3.5~km (although larger depths are also possible), we obtain a best-fit
age of $\sim0.95$ and $\sim1.3$~Ga for population P1 and P2,
respectively. \\

This latter result is puzzling, since such a large crater is not
expected to be so young. Note that the inferred age is quite
insensitive to the adopted scaling law or impactor population.  We
estimated that the impactor that formed Massalia was in the size range
7-9~km \citep{sie11}. The current frequency of such impacts is about
one every 9~Ga.  The computed a priori probability that such an event
happened in the last $1$~Ga, is $\sim11\%$. On the other hand, knowing
that the Massalia event did happen within the last 4.5~Ga, the
probability that such event occurred in the last 1~Ga, is
$\sim25\%$. These numbers apply for the present main belt impact rate,
and thus certainly represent an upper limit because it is believed
that the impact rate in the primordial main belt was at least a factor
$\sim2-4$ more intense than today \citep{mor10}.  This would imply
  that Massalia event more likely happened early on rather than
  recently.  Thus, in conclusion, it is likely that other processes
  may be responsible for a lack of craters within Narbonensis
  \citep{mas11,nic11}.\\

As discussed in the previous section, this unit has relatively high
topographical slopes and episodes of slopes slumping may have induced
crater erasing (see Fig. \ref{slopes}, upper panels). The presence of
significant rims slumping is supported by the V-shaped topographical
profile of the crater \citep{pre11}.  Comparing the observed profile
with a typical profile of a fresh crater \citep{cre11}, we derive that
several 100s~m of rim material may have been displaced toward the
center of the crater, which may be enough to explain the relatively
young age of this unit.

\section{Discussions and conclusions}

The main result of our crater retention age analysis is that it
confirms a prolonged and complex collisional evolution of Lutetia.  As
shown for previous asteroids visited by spacecraft, collisions play a
major role in the evolution of any asteroid, being largely responsible
of their shapes, internal structure and geomorpohological
features. The latter play also an important role for the
understanding of surface specrophotometric properties.\\

All these collisional-related processes are well documented on
Lutetia, and can be used to constrain its evolution.  The derived ages
of the main units of Lutetia show its active collisional history,
lasting for about 4~Ga.  The extremely young Bt1a unit, with an
age of $<220$~Ma, indicates that major (collisional) events occurred
until very recent times.  We also find evidence on the oldest Achaia
region -and possibly Noricum region-, of a non-uniform radial strength
profile, possibly due to the effects of previous collisions that
produced a highly fractured surface on top a competent interior.  In
this respect, Lutetia resembles what has been found on other much
larger bodies like Mercury and the Moon \citep{mar11}.  It is also
  noteworthy that the observed cratering seems to be produced by a
  population having a shallower cumulative slope than predicted by the
  \cite{bot05a} model.  The overall slope seems to be consistent with
  recent observations \citep{gla09}, although our size range of
  interest extends beyong their obsevational limits. This result, if
  confirmed by further studies, will require a revision of the present
  collisional models.\\

On the other hand, according to the present theories of main belt
evolution, Lutetia should be a primordial object \citep{bot05a}. This
is also confirmed by Lutetia's high density that makes it unlikely to
be a fragment of a larger body \citep{wei11}.  This consideration is,
however, in contradiction with the derived crater retention ages.
Either the chronology scheme is not accurate, or some major event
occurred in Lutetia history to reset its surface.\\

  Concerning the adopted lunar chronology, it likely underestimates
  the real ages of main belt asteroids. Indeed, the exponential
  increase of the lunar impactor flux for ages older than 3.5~Ga did
  not likely take place in the main belt unless the main belt suffered
  an intense cometary bombardment. In fact, in the current scenario of
  main belt evolution during the Late Heavy Bombardment (LHB) at $\sim
  3.9$~Ga, the main belt depleted by a factor of 2-4 at most
  \citep{mor10}.  On the other hand, the lunar chronology
  (Equ.~\ref{chrono_eq}) predicts an increase in the impactor flux of
  about a factor of $\sim5$ and $\sim40$ in the time spans
  $3.5-3.9$~Ga and $3.5-4.2$~Ga, respectively. This steep
    increase in the impactor flux results in too young crater
    retention ages of asteroid surfaces.  However, dynamical models of
    the early evolution of the main belt are not yet robust enough to
    successfully be used for precise age determination.\\

Concerning possible resetting event(s), the most energetic event that
we can infer is the formation of Massalia crater, which, according to
the previous discussion, was not able to reset the whole
surface. Unless this conclusion is affected by poorly constrained
  parameters or other more energetic event(s) took place in the
  Lutetia southern hemisphere (not imaged by OSIRIS), this option
  appears untenable. \\

Lutetia's crater age conundrum still remains unsolved. Nevertheless,
we expect major improvements in our theoretical and observational
understandings of the main belt in the near future.  In particular,
the Dawn mission arrived at Vesta, the second largest asteroid, in
July 2011. High resolution imaging of Vesta will help to constrain the
early impact history of the main belt and the evolution of its
primordial asteroids, Lutetia included.\\

{\bf Acknowledgments}

We thank the referee D.~O'Brien and an anonymous referee for helpful
comments which improved the manuscript.\\ 
OSIRIS was built by a
consortium of the Max-Planck-Institut f{\"u}r Sonnensystemforschung,
Katlenburg-Lindau, Germany, CISAS - University of Padova, Italy, the
Laboratoire d'Astrophysique de Marseille, France, the Instituto de
Astrofísica de Andalucia, CSIC, Granada, Spain, the Research and
Scientific Support Department of the European Space Agency, Noordwijk,
The Netherlands, the Instituto Nacional de Tecnica Aeroespacial,
Madrid, Spain, the Universidad Politechnica de Madrid, Spain, the
Department of Physics and Astronomy of Uppsala University, Sweden, and
the Institut f{\"u}r Datentechnik und Kommunikationsnetze der
Technischen Universit{\"a}t Braunschweig, Germany.
The support of the national funding agencies of Germany (DLR), France
(CNES), Italy (ASI), Spain (MEC), Sweden (SNSB), and the ESA Technical
Directorate is gratefully acknowledged.\\
We thank the Rosetta Science Operations Centre and the Rosetta Mission
Operations Centre for the successful flyby of (21) Lutetia.


\newpage

\end{document}